\documentclass{article}

\usepackage{microtype}
\usepackage{graphicx}
\usepackage{subfigure}
\usepackage{booktabs} 
\usepackage{amsmath}
\usepackage{amssymb}
\usepackage{algorithm}
\usepackage{algorithmic}
\usepackage{multirow}

\usepackage{hyperref}

\usepackage[accepted]{mlsys2025}

\mlsystitlerunning{StreamServe: Adaptive Speculative Flows for Low-Latency Disaggregated LLM Serving}

\begin{document}

\twocolumn[
\mlsystitle{StreamServe: Adaptive Speculative Flows for Low-Latency Disaggregated LLM Serving}




\begin{mlsysauthorlist}
\mlsysauthor{Satyam Kumar *}{}
\mlsysauthor{Arpit Singh Gautam *}{}
\mlsysauthor{Kailash Talreja}{}
\mlsysauthor{Saurabh Jha}{}
\end{mlsysauthorlist}

\mlsyscorrespondingauthor{Satyam Kumar}{satyamkumar9742@gmail.com}
\mlsyscorrespondingauthor{Arpit Singh Gautam}{arpitsinghgautam777@gmail.com}

\mlsyskeywords{Machine Learning, MLSys, Large Language Models, LLM Serving, Disaggregated Inference, Speculative Decoding, Adaptive Routing, Request Scheduling, GPU Utilization, Latency Optimization}
\vskip 0.3in

\begin{abstract}
Efficient large language model (LLM) serving requires navigating inherent trade-offs between throughput and latency across diverse, bursty workloads. We present \textbf{StreamServe}, a disaggregated prefill--decode serving architecture that integrates (i) \emph{metric-aware routing} across compute lanes with (ii) \emph{adaptive speculative decoding} that tunes speculation depth online based on runtime signals. StreamServe comprises four tightly coupled components: \textbf{StreamScheduler} (request orchestration), \textbf{FlowGuard} (multi-signal routing), \textbf{PipeServe-Engine} (disaggregated prefill/decode execution on multi-GPU), and \textbf{SpecuStream} (runtime-adaptive speculation).

We evaluate StreamServe on four benchmarks (ALPACA, GSM8K, HUMANEVAL, SUM; 80 queries each, 320 total) using \textbf{4$\times$~A800-40GB} GPUs configured as two stream pairs. Across all evaluated workloads, we observe systematic latency reductions of 11--18$\times$ relative to tensor-parallel vLLM baselines, with throughput reaching up to 2235 tokens/s on summarization tasks. Time-per-output-token (TPOT) remains stable across configurations, suggesting that the observed gains arise from architectural efficiency rather than token-quality degradation. While the current evaluation is limited to a single-node 4-GPU configuration, these findings indicate that the joint adaptation of routing and speculation within a disaggregated framework constitutes a qualitatively distinct operating regime for LLM inference.

\end{abstract}
]
\footnotetext{* Equal contribution.}



\section{Introduction}
\label{submission}

Large language models underpin a growing range of applications spanning natural language understanding, code generation, reasoning, and content synthesis. Deploying these models in production settings introduces fundamental systems challenges: workloads are heterogeneous and bursty, and the interaction between throughput and per-request latency is governed by architectural choices that are difficult to optimize in isolation. Monolithic serving architectures couple prefill and decode phases, resulting in resource contention and suboptimal hardware utilization~\cite{Kwonetal2023}. A body of recent work addresses individual aspects of this challenge---continuous batching~\cite{Chen2024ContinuousBatching}, memory-efficient attention via PagedAttention~\cite{Kwonetal2023}, and speculative decoding~\cite{Leviathan2023SpeculativeDecoding, Li2024EAGLE}---while disaggregated serving separates prefill and decode onto dedicated resources~\cite{Zhong2024DistServe,Agrawal2024Sarathi}, yielding measurable gains through chunked-prefill and stall-free scheduling~\cite{Agrawal2024Sarathi}. Speculative systems such as EAGLE and Medusa further reduce latency by verifying multiple predicted tokens in parallel~\cite{Li2024EAGLE2, Cai2024Medusa}.

However, we identify two gaps in existing systems: (i) disaggregated architectures generally lack \emph{metric-aware routing} that adapts to real-time system state across compute lanes, and (ii) speculative decoding methods employ \emph{fixed} speculation depths that do not respond to dynamic load and acceptance patterns. We hypothesize that jointly adapting routing and speculation within a disaggregated framework can yield compounding efficiency gains that neither mechanism achieves independently.

\textbf{Contributions.} Building upon the prefill--decode disaggregation paradigm established by DistServe~\cite{Zhong2024DistServe} and Sarathi-Serve~\cite{Agrawal2024Sarathi}, we introduce the following components:
\begin{enumerate}
    \item \emph{FlowGuard}, a multi-signal metric-aware router that combines cache-reuse, memory utilization, queue depth, and active load signals to inform request routing across disaggregated compute lanes.
    \item \emph{SpecuStream}, an adaptive speculative decoding mechanism that dynamically adjusts speculation depth at runtime based on acceptance rate gradients, system load, and throughput targets.
    \item \emph{Joint optimization}: The systematic co-adaptation of routing and speculation within a disaggregated serving framework. We find that this integration yields performance improvements that are super-additive relative to the individual components.
    \item An empirical evaluation on four benchmarks (320 queries total) using 4$\times$A800 GPUs, with detailed ablations, percentile analysis, and concurrency scaling characterization.
\end{enumerate}

We note that the disaggregated prefill--decode architecture itself is \emph{not} our contribution; our contributions lie in the routing and adaptive speculation mechanisms built atop this architectural foundation, and in the empirical characterization of their interaction.

\section{Related Work}

\subsection{Disaggregated LLM Serving Architectures}

A growing line of work investigates disaggregating the prefill and decode phases of LLM inference to mitigate phase-specific bottlenecks. DistServe~\cite{Zhong2024DistServe} introduced this paradigm by separating prefill computation from decode stages onto dedicated hardware, demonstrating that such decoupling reduces resource contention and improves goodput relative to monolithic designs. Sarathi-Serve~\cite{Agrawal2024Sarathi} extends this foundation with chunked-prefilling and stall-free batch scheduling, reporting 2.6--5.6$\times$ improvements in throughput-latency trade-offs over conventional vLLM deployments. Seesaw~\cite{Su2025Seesaw} further develops disaggregated architectures through dynamic model re-sharding strategies that adapt resource allocation to workload characteristics.

\textbf{Comparison with Mooncake.} Mooncake~\cite{Qin2024Mooncake} proposes a KVCache-centric disaggregated architecture emphasizing prefix-cache-aware routing for multi-turn conversations. While Mooncake optimizes primarily for cache locality through KV reuse from shared prefixes, FlowGuard differs in two respects: (1) \emph{Multi-signal scoring}: FlowGuard integrates four complementary signals (cache hit rate, memory utilization, queue depth, and active load) rather than optimizing for cache reuse alone---a distinction that becomes important for mixed workloads where prefix sharing is limited; (2) \emph{Overload-aware routing}: FlowGuard incorporates dynamic overload detection that excludes saturated workers from routing consideration, a mechanism absent from Mooncake's scheduler.

These works establish disaggregation as a foundational architectural principle for efficient LLM serving. However, existing disaggregated systems generally rely on static scheduling policies that do not adapt to real-time system metrics or dynamic workload variations, leaving potential efficiency gains unrealized.

\subsection{Speculative Decoding and Token Prediction}

Speculative decoding accelerates LLM inference by generating multiple candidate tokens in parallel and verifying them against the target model. The foundational formulation by~\cite{Leviathan2023SpeculativeDecoding} establishes that speculative token generation and parallel verification can reduce end-to-end latency while preserving output distribution. Subsequent work has explored increasingly sophisticated speculation strategies. EAGLE~\cite{Li2024EAGLE} reconsiders feature uncertainty in draft token generation, while EAGLE-2~\cite{Li2024EAGLE2} introduces dynamic draft tree speculation that adapts tree structure to model-specific patterns. EAGLE-3~\cite{Li2025EAGLE3} addresses practical deployment challenges when scaling speculative decoding to larger models on high-end GPUs. Medusa~\cite{Cai2024Medusa} takes a complementary approach with multiple decoding heads that predict several future tokens simultaneously. Recent work on speculative verification~\cite{Kim2025SpeculativeVerification} incorporates information gain metrics to refine candidate token selection.

\textbf{Comparison with Adaptive Speculation Methods.} Table~\ref{tab:speculation_comparison} positions SpecuStream relative to recent adaptive speculation approaches. SpecDec++~\cite{Huang2024SpecDecPP} adapts candidate lengths based on token-level prediction confidence at per-token granularity. AdaServe~\cite{Li2024AdaServe} customizes speculation depth according to SLO targets, enabling per-request adaptation. SpecuStream differs from both in that it incorporates \emph{system-level signals}---load factor, recent throughput, and acceptance rate gradients---alongside token acceptance rates. This design enables co-optimization with routing decisions, as FlowGuard and SpecuStream share the same metric infrastructure. While SpecDec++ and AdaServe operate independently of request routing, SpecuStream's system-aware formulation enables coordinated optimization across the serving stack.

\begin{table}[h!]
\centering
\resizebox{\columnwidth}{!}{%
\begin{tabular}{lccc}
\toprule
\textbf{Approach} & \textbf{Adaptation Signal} & \textbf{Scope} & \textbf{Routing-Aware} \\
\midrule
SpecDec++ & Token confidence & Per-token & No \\
AdaServe & SLO targets & Per-request & No \\
\textbf{SpecuStream} & \textbf{Accept. gradients + load + tput} & \textbf{Per-request} & \textbf{Yes} \\
\bottomrule
\end{tabular}%
}
\caption{Comparison of adaptive speculation approaches. SpecuStream incorporates system-level signals and co-optimizes with routing decisions.}
\label{tab:speculation_comparison}
\end{table}

Despite these advances, existing speculative decoding methods generally employ fixed speculation depths that do not respond to changing system conditions or request characteristics, potentially forgoing efficiency gains during periods of variable resource availability.

\subsection{Intelligent Request Scheduling and Routing}

Request scheduling directly shapes both throughput and latency in LLM serving systems. \cite{Zheng2024LearningToRank} develop learning-to-rank approaches that prioritize request ordering based on predicted completion times. \cite{Li2025ThroughputOptimalScheduling} provide theoretical frameworks for throughput-optimal request ordering, while \cite{Bari2025OptimalScheduling} establish rigorous foundations for scheduling algorithm design. FairBatching~\cite{Chen2025FairBatching} introduces fairness-aware batch formation that balances performance across diverse request types. FluidGuided scheduling~\cite{Ao2025FluidGuided} applies flow-based models for online scheduling that adapt to resource availability. However, most scheduling approaches operate at the request-ordering level without incorporating multi-signal metrics for routing across multiple compute lanes, and none systematically integrate scheduling with adaptive speculation.

\subsection{Memory Optimization and KV-Cache Management}

Efficient memory management is a binding constraint in LLM inference. PagedAttention~\cite{Kwonetal2023} applies virtual memory techniques to KV-cache management, reducing memory fragmentation and enabling larger batch sizes---a technique now standard across modern serving systems. Prompt caching~\cite{Zhang2024PromptCache} enables reuse of cached KV values across requests with shared prefixes, improving efficiency for multi-turn and structured workloads. AttentionStore~\cite{Ramaseshaan2025AttentionStore} extends this with cost-effective attention reuse for multi-turn conversations. FlashInfer~\cite{Chen2024FlashInfer} develops efficient attention kernels optimized for GPU execution. At the infrastructure level, NIXL~\cite{Wei2025NIXL} provides high-performance peer-to-peer KV-cache transfer between distributed GPU workers, and LMCache~\cite{Dey2025LMCache} introduces enterprise-scale cache management with eviction and pinning strategies. While these techniques have individually advanced efficiency, their systematic integration with disaggregated architectures and adaptive routing remains an open problem.

\subsection{Batching and Continuous Processing}

Continuous batching is a foundational optimization for LLM inference. \cite{Chen2024ContinuousBatching} characterize continuous batching strategies that dynamically form request batches as arrivals and completions occur, improving hardware utilization over static batching. Lookahead decoding~\cite{Dey2024Lookahead} breaks sequential dependencies in token generation through lookahead mechanisms. Dynamic batch budget allocation~\cite{Choi2025AdaptiveOutput} adapts batch sizes to system constraints and predicted completion times. Work on head-of-line blocking~\cite{Fu2024HeadOfLine} addresses inefficiencies from request latency heterogeneity in batched systems. These innovations improve aggregate throughput but can sacrifice individual request latency, particularly for latency-sensitive requests mixed with throughput-oriented batch processing.

\subsection{Motivation for StreamServe}

While individual components of the LLM inference stack have seen substantial progress, the joint optimization of disaggregated architectures with metric-aware routing and adaptive speculative decoding remains unexplored. Existing systems either focus on disaggregation without intelligent routing, or employ sophisticated speculation strategies without adapting to real-time system dynamics. This gap motivates StreamServe: a unified framework that co-adapts scheduling, routing, and speculation based on dynamic system state and workload characteristics. Our central hypothesis is that these components interact in a manner that produces qualitatively different performance regimes when optimized jointly, rather than yielding merely additive improvements.

\section{Methodology}

StreamServe implements a disaggregated LLM serving architecture comprising four tightly integrated components: StreamScheduler for request orchestration, FlowGuard for metric-aware routing, PipeServe-Engine for disaggregated prefill-decode execution, and SpecuStream for adaptive speculation.

\subsection{Notation}

Table~\ref{tab:notation} summarizes the key symbols used throughout this paper.

\begin{table}[h!]
\centering
\resizebox{\columnwidth}{!}{%
\begin{tabular}{cl}
\toprule
\textbf{Symbol} & \textbf{Definition} \\
\midrule
$N$ & Number of stream pairs \\
$w, i$ & Worker/stream pair index \\
$C_w$ & Cache hit rate for worker $w$ \\
$M_w \in [0,1]$ & Normalized memory utilization \\
$Q_w \in [0,1]$ & Normalized queue depth \\
$L_w \in [0,1]$ & Normalized active request load \\
$\alpha_1, \ldots, \alpha_4$ & Routing weights ($\sum_j \alpha_j = 1$) \\
$S_w$ & FlowGuard routing score for worker $w$ \\
$\omega_w$ & Overload score for worker $w$ \\
$\tau$ & Overload threshold \\
\midrule
$\mathbf{f} \in \mathbb{R}^h$ & Flow vector tracking acceptance gradients \\
$h$ & Flow vector history length (default: 10) \\
$a_t$ & Acceptance rate at time $t$ \\
$\delta_t$ & Acceptance rate gradient at time $t$ \\
$\mathcal{M}_f$ & Flow magnitude (gradient volatility) \\
$d_{\text{spec}}$ & Computed speculation depth \\
$d_{\text{base}}$ & Baseline speculation depth (default: 5) \\
$\gamma$ & Amplification factor (default: 5) \\
$\phi_{\text{load}}$ & Load adaptation factor \\
$\phi_{\text{tput}}$ & Throughput scaling factor \\
$\tau_{\text{target}}$ & Target throughput (tokens/second) \\
$\tau_{\text{recent}}$ & Recently observed throughput \\
\midrule
$\ell_p$ & Prompt length (tokens) \\
$\ell_g$ & Number of generated tokens \\
$d_{\text{model}}$ & Model hidden dimension \\
\bottomrule
\end{tabular}%
}
\caption{Summary of notation used throughout this paper.}
\label{tab:notation}
\end{table}

\subsection{System Architecture Overview}

\begin{figure*}[ht]
    \centering
    \includegraphics[width=0.9\linewidth]{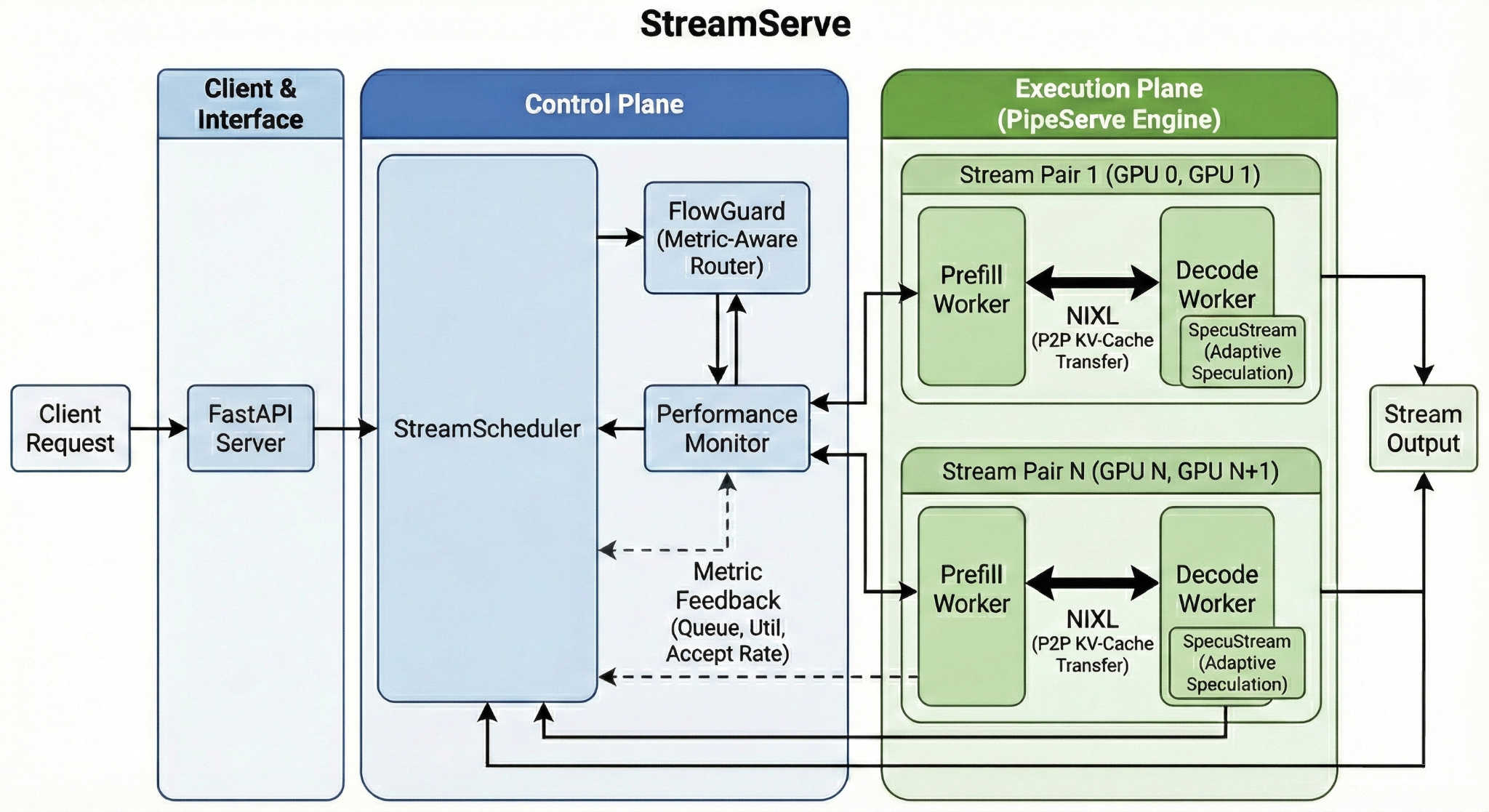}
    \caption{\textbf{StreamServe Architecture.} The \textbf{Control Plane} orchestrates requests via the \textbf{StreamScheduler}, which queries the \textbf{FlowGuard} router for optimal placement. The router relies on real-time feedback (dashed lines) from the \textbf{Performance Monitor} regarding queue depth and speculation rates. The \textbf{Execution Plane} (PipeServe Engine) runs disaggregated prefill and decode phases on separate GPUs, linked by \textbf{NIXL} for high-bandwidth P2P KV-cache transfer.}
    \label{fig:system_arch}
\end{figure*}

StreamServe deploys \( N \) stream pairs across \( 2N \) GPUs, where each stream pair \( i \in \{1, \ldots, N\} \) consists of a prefill worker on GPU \( 2i \) and a decode worker on GPU \( 2i+1 \). Algorithm~\ref{alg:streamserve_full} describes the system architecture. Given an incoming request \( r \) with prompt \( p \) and generation parameters \( \theta \), the FastAPI server tokenizes the prompt to obtain token identifiers \( \textit{token\_ids} \), constructs an internal request representation \( req \), and invokes the StreamScheduler. The scheduler routes \( req \) through FlowGuard to select the target stream pair, enqueues to the selected prefill queue \( Q_{\text{prefill}_i} \), and coordinates subsequent decode processing.

\begin{algorithm}[tb]
   \caption{StreamServe Full Architecture}
   \label{alg:streamserve_full}
\begin{algorithmic}
   \STATE {\bfseries Input:} Request $r$ with prompt $p$, params $\theta$
   \STATE Init server $S$, manager $M$, queues $Q_P, Q_D$, scheduler $Sch$
   \FOR{stream pair $i$ in $1$ to $N$}
   \STATE Launch prefill $P_i$ on GPU $2i$, decode $D_i$ on GPU $2i+1$
   \ENDFOR
   \WHILE{$S$ running}
   \STATE Receive $r$; tokenize $p \to toks$; create $req$
   \STATE $Sch$.route($req$) via FlowGuard $\to Q_{P_i}$
   \STATE $P_i$: prefill $toks$, transfer KV via NIXL $\to Q_{D_i}$
   \STATE $D_i$: decode with SpecuStream, report metrics
   \STATE Return response to client
   \ENDWHILE
   \STATE Shutdown scheduler, queues, processes
\end{algorithmic}
\end{algorithm}

\subsection{FlowGuard: Metric-Aware Routing}

FlowGuard implements a multi-factor scoring function to select the target compute lane for each incoming request. The design principle is to route requests toward workers that maximize cache reuse while avoiding resource saturation. For stream pair \( w \), the score function is defined as:

\begin{equation}
S_w = \alpha_1 C_w + \alpha_2(1 - M_w) + \alpha_3(1 - Q_w) + \alpha_4(1 - L_w)
\label{eq:flowguard_score}
\end{equation}

where \( C_w \) denotes the cache hit rate, \( M_w \in [0,1] \) is normalized memory utilization, \( Q_w \in [0,1] \) is normalized queue depth, \( L_w \in [0,1] \) indicates normalized active request load, and \( \alpha_1, \alpha_2, \alpha_3, \alpha_4 \) are tunable weights satisfying \( \sum_{j=1}^4 \alpha_j = 1 \). 

The score function rewards workers with high cache reuse (\( C_w \)) while penalizing high memory pressure, deep queues, and heavy load. Higher scores correspond to more favorable routing targets. In our experiments, we set \( \alpha_1 = 0.4 \), \( \alpha_2 = 0.1 \), \( \alpha_3 = 0.3 \), \( \alpha_4 = 0.2 \). These weights reflect our empirical finding that cache locality and queue depth most directly govern per-request latency in disaggregated configurations.

FlowGuard additionally implements dynamic overload detection via a composite threshold mechanism:

\begin{equation}
\text{Overload}(w) = \begin{cases}
\text{True} & \text{if } \omega_w > \tau \\
\text{False} & \text{otherwise}
\end{cases}
\label{eq:overload_detection}
\end{equation}

where the overload score \( \omega_w \) is computed as:

\begin{equation}
\omega_w = \frac{M_w}{100} + 2 \cdot \frac{Q_w}{Q_{\max}}
\label{eq:overload_score}
\end{equation}

The overload score assigns greater weight to queue depth (factor of 2) than to memory utilization, reflecting the observation that queueing delays have more immediate impact on per-request latency than memory pressure. The threshold \( \tau = 0.85 \) was empirically selected to balance load distribution with routing stability. Workers exceeding this threshold are excluded from routing consideration. When all workers are saturated, the system falls back to the worker with minimum queue depth:

\begin{equation}
w^* = \arg\min_{i \in \{1,\ldots,N\}} Q_i
\label{eq:fallback_selection}
\end{equation}

Algorithm~\ref{alg:flowguard} details the complete selection procedure.

\begin{algorithm}[tb]
   \caption{FlowGuard Worker Selection}
   \label{alg:flowguard}
\begin{algorithmic}
   \STATE {\bfseries Input:} Metrics $\{m_i\}_{i=1}^N$, weights $W$, thresholds $T$
   \STATE Collect metrics: $\forall i$: $perf_i, load_i \gets$ fresh values
   \STATE $load_i.qd \gets Q_{P_i}.\text{size}()$ \COMMENT{queue depth}
   \STATE $avail \gets \emptyset$, $scores \gets \{\}$
   \FOR{each $m_i$}
   \IF{$m_i$ not stale \AND not overloaded$(m_i, T)$}
   \STATE $scores[i] \gets \text{calc\_score}(m_i, W)$
   \STATE $avail.\text{add}(i)$
   \ENDIF
   \ENDFOR
   \IF{$avail = \emptyset$}
   \STATE \textbf{return} $\arg\min_i load_i.qd$ \COMMENT{fallback}
   \ELSE
   \STATE \textbf{return} $\arg\max_{i \in avail} scores[i]$
   \ENDIF
\end{algorithmic}
\end{algorithm}

\subsection{PipeServe-Engine: Disaggregated Execution}

Each stream pair executes prefill and decode phases on dedicated GPU resources with NIXL-based peer-to-peer KV-cache transfer. For a request with prompt length \( \ell_p \) tokens, the prefill worker computes:

\begin{equation}
\text{KV}_{\text{prefill}} = \text{Attention}(Q_p, K_p, V_p) \in \mathbb{R}^{\ell_p \times d_{\text{model}}}
\label{eq:prefill_computation}
\end{equation}

where \( Q_p, K_p, V_p \) represent query, key, and value projections for the prompt tokens, and \( d_{\text{model}} \) denotes the model hidden dimension. The computed KV-cache is transferred via NIXL's GPU-direct P2P protocol to the decode worker with bandwidth:

\begin{equation}
B_{\text{transfer}} = \frac{|\text{KV}_{\text{prefill}}| \cdot \text{sizeof}(\text{float16})}{t_{\text{transfer}}}
\label{eq:transfer_bandwidth}
\end{equation}

The decode worker then generates \( \ell_g \) output tokens autoregressively:

\begin{equation}
\text{KV}_{\text{total}} = \text{Concat}(\text{KV}_{\text{prefill}}, \text{KV}_{\text{decode}})
\label{eq:kv_concatenation}
\end{equation}

Algorithm~\ref{alg:pipeserve} describes the complete pipeline operation.

\begin{algorithm}[tb]
   \caption{PipeServe Engine Stream Pair}
   \label{alg:pipeserve}
\begin{algorithmic}
   \STATE {\bfseries Input:} Stream $i$, queues $Q_{P_i}, Q_{D_i}$, model $M$, config $C$
   \STATE Init $llm_P$ on GPU $2i$ (NIXL producer)
   \STATE Init $llm_D$ on GPU $2i{+}1$ (NIXL consumer + spec)
   \WHILE{running}
   \STATE $req \gets Q_{P_i}.\text{get}()$
   \IF{$req = $ None} \STATE \textbf{break} \ENDIF
   \STATE $kv \gets llm_P.\text{prefill}(req.prompt)$
   \STATE Transfer $kv$ via NIXL P2P to decode
   \STATE $Q_{D_i}.\text{put}(req)$
   \STATE $req \gets Q_{D_i}.\text{get}(timeout{=}0.2)$
   \IF{$req.error$}
   \STATE Propagate error
   \ELSE
   \STATE Adapt spec depth via SpecuStream
   \STATE $out \gets llm_D.\text{generate}(req)$
   \STATE Compute \& report metrics
   \ENDIF
   \ENDWHILE
   \STATE Shutdown NIXL, LLM instances
\end{algorithmic}
\end{algorithm}

\subsection{SpecuStream: Adaptive Speculation Optimization}

SpecuStream dynamically adjusts speculative decoding depth based on real-time system metrics. The key observation motivating this design is that the relationship between speculation depth and effective throughput is non-stationary: it depends jointly on model-specific acceptance patterns and system-level resource availability, both of which vary over time.

The algorithm maintains a flow vector \( \mathbf{f} \in \mathbb{R}^{h} \) of length \( h \) (typically \( h=10 \)) tracking historical acceptance rate gradients. Given current acceptance rate \( a_t \) at time \( t \), we compute:

\begin{equation}
\delta_t = a_t - \frac{1}{h}\sum_{j=0}^{h-1} f_j
\label{eq:acceptance_gradient}
\end{equation}

The gradient \( \delta_t \) captures how the current acceptance rate deviates from the recent average. Positive gradients indicate improving speculation accuracy; negative gradients suggest degradation, signaling a shift in workload characteristics.

The flow vector is updated circularly: \( f_{t \bmod h} \leftarrow \delta_t \). We then compute flow magnitude as:

\begin{equation}
\mathcal{M}_f = \frac{1}{h}\sum_{j=0}^{h-1} |f_j|
\label{eq:flow_magnitude}
\end{equation}

The flow magnitude \( \mathcal{M}_f \) quantifies the volatility of acceptance rates over recent history. We find empirically that high volatility correlates with changing workload patterns, warranting conservative speculation, while stable high acceptance enables deeper speculation.

To incorporate throughput objectives, we define a scaling factor:

\begin{equation}
\phi_{\text{tput}} = \max\left(1, \frac{\tau_{\text{target}}}{\max(\tau_{\text{recent}}, 1)}\right)
\label{eq:throughput_scaling}
\end{equation}

where \( \tau_{\text{target}} \) represents the target throughput (e.g., 400 tokens/second) and \( \tau_{\text{recent}} \) denotes recently observed throughput. When throughput falls below the target, \( \phi_{\text{tput}} > 1 \) encourages deeper speculation to increase token generation rate; when throughput meets or exceeds the target, \( \phi_{\text{tput}} = 1 \) maintains the current depth.

Load adaptation is computed as:

\begin{equation}
\phi_{\text{load}} = 1 - \min(l_w, 0.9)
\label{eq:load_adaptation}
\end{equation}

where \( l_w \) represents the normalized load factor for worker \( w \). Under high load (\( l_w \to 0.9 \)), \( \phi_{\text{load}} \to 0.1 \), reducing speculation depth to conserve resources. Under low load, \( \phi_{\text{load}} \to 1 \), permitting deeper speculation.

The speculative depth is then computed as:

\begin{equation}
d_{\text{spec}} = d_{\text{base}} + (a_t \cdot \mathcal{M}_f \cdot \gamma) \cdot \phi_{\text{load}} \cdot \phi_{\text{tput}}
\label{eq:optimal_depth}
\end{equation}

where \( d_{\text{base}} \) is the baseline speculation depth (typically 5) and \( \gamma \) is an amplification factor (typically \( \gamma = 5 \)). The final depth is clipped to a valid range:

\begin{equation}
d_{\text{spec}}^* = \text{clip}(d_{\text{spec}}, d_{\min}, d_{\max})
\label{eq:depth_clipping}
\end{equation}

where typically \( d_{\min} = 2 \) and \( d_{\max} = 20 \). Concurrently, micro-batch size is adjusted inversely to speculation depth:

\begin{equation}
b_{\text{micro}} = \max\left(1, \left\lfloor \frac{16 \cdot 5}{d_{\text{spec}}^*} \right\rfloor\right)
\label{eq:microbatch_size}
\end{equation}

This inverse relationship ensures that deeper speculation---which requires more memory and computation per sequence---is accompanied by smaller batch sizes to maintain latency constraints and avoid memory pressure.

Finally, projected throughput is estimated using exponential smoothing:

\begin{equation}
\tau_{\text{proj}} = \tau_{\text{recent}} \cdot (1 + a_t \cdot 0.5)
\label{eq:projected_throughput}
\end{equation}

\begin{equation}
\tau_{\text{recent}}^{\text{new}} = 0.9 \cdot \tau_{\text{recent}} + 0.1 \cdot \tau_{\text{proj}}
\label{eq_exponential_smoothing}
\end{equation}

Algorithm~\ref{alg:specustream} provides the complete adaptation procedure. These adaptations occur per request based on the selected worker's current metrics, forming a closed-loop control mechanism that responds to system state fluctuations.

\begin{algorithm}[tb]
   \caption{SpecuStream Adaptation}
   \label{alg:specustream}
\begin{algorithmic}
   \STATE {\bfseries Input:} Metrics $m$: accept rate $a$, load $l$, throughput $t$
   \STATE $\delta \gets a - \text{mean}(\mathbf{f})$; \quad $\mathbf{f}[idx] \gets \delta$
   \STATE $idx \gets (idx + 1) \mod h$
   \STATE $mag \gets \text{mean}(|\mathbf{f}|)$
   \STATE $scale \gets \max(1, \tau_{target} / \max(t, 1))$
   \STATE $adj \gets 1 - \min(l, 0.9)$
   \STATE $d \gets d_{base} + (a \cdot mag \cdot \gamma) \cdot adj \cdot scale$
   \STATE $d^* \gets \text{clip}(d, d_{min}, d_{max})$
   \STATE $b_{micro} \gets \max(1, \lfloor 16 \cdot 5 / d^* \rfloor)$
   \STATE $t_{proj} \gets t \cdot (1 + a \cdot 0.5)$
   \STATE $\tau_{recent} \gets 0.9 \cdot \tau_{recent} + 0.1 \cdot t_{proj}$
   \STATE {\bfseries Output:} $\{d^*, b_{micro}, t_{proj}\}$
\end{algorithmic}
\end{algorithm}

\subsection{Performance Metrics}

StreamServe collects system metrics at 500-millisecond intervals. For each completed request, end-to-end latency is defined as:

\begin{equation}
\text{Latency} = t_{\text{end}} - t_{\text{start}}
\label{eq:latency_computation}
\end{equation}

Time per output token (TPOT) is calculated as:

\begin{equation}
\text{TPOT} = \frac{\sum_{k=1}^{\ell_g} (t_k - t_{k-1})}{\ell_g}
\label{eq:tpot_computation}
\end{equation}

Throughput is defined as total tokens processed divided by latency:

\begin{equation}
\text{Throughput} = \frac{\ell_p + \ell_g}{\text{Latency}}
\label{eq:throughput_computation}
\end{equation}

where \( t_k \) denotes the generation time for the \( k \)-th output token, \( \ell_p \) is prompt length, and \( \ell_g \) is the number of generated tokens. These metrics inform both API responses and subsequent FlowGuard routing decisions, establishing a closed-loop system in which all components---scheduling, routing, and speculation---adapt continuously to observed workload characteristics and system state.

%
%

\begin{figure*}[t!]
\centering
\subfigure[Latency (s) Comparison]{\includegraphics[width=0.32\textwidth]{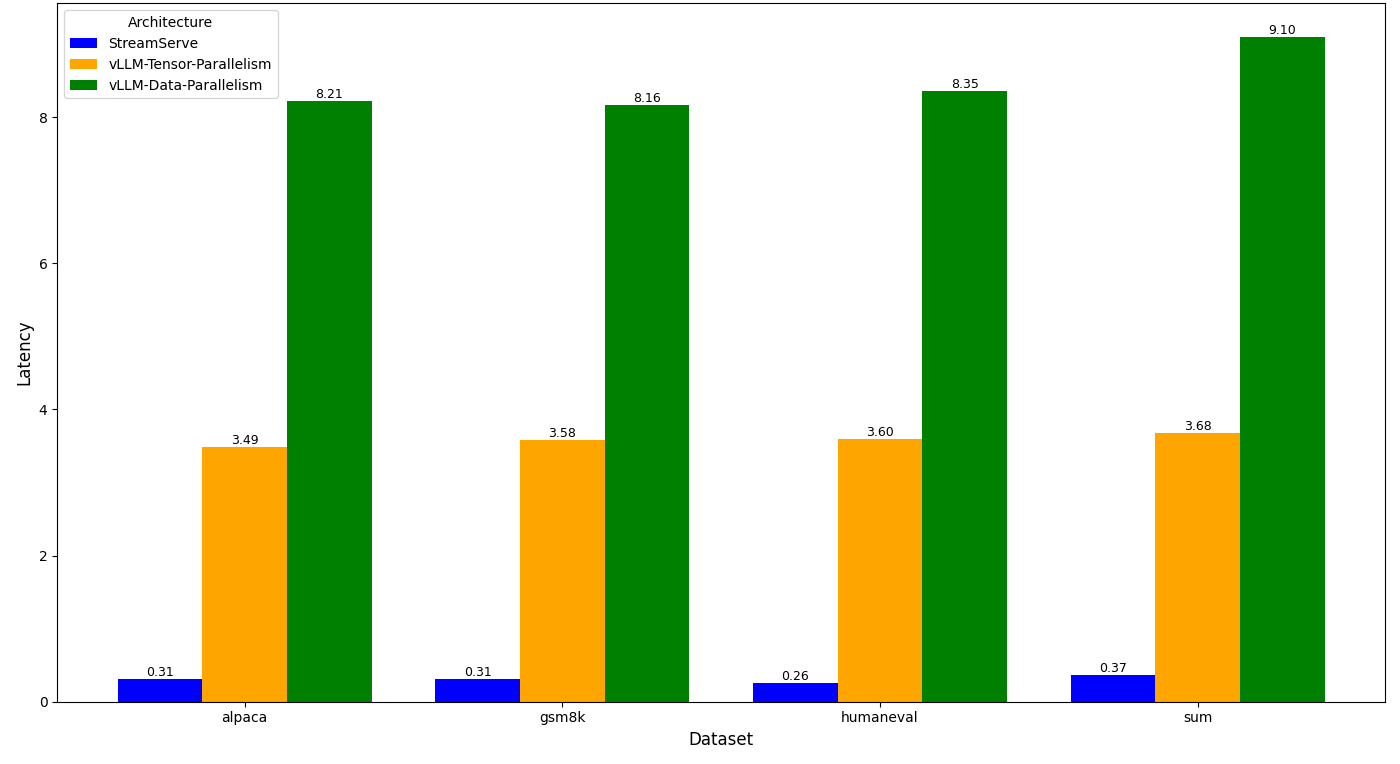}}
\hfill
\subfigure[Throughput (tokens/s) Comparison]{\includegraphics[width=0.32\textwidth]{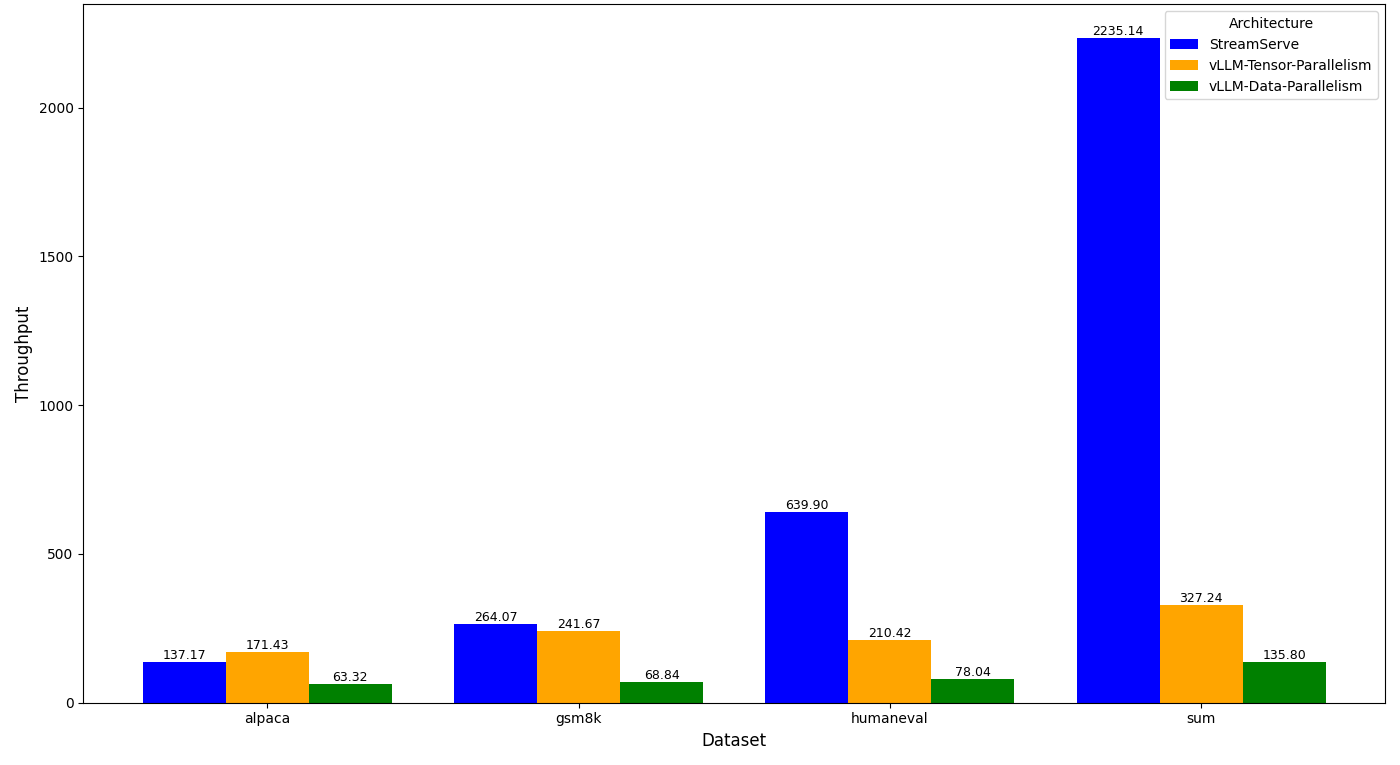}}
\hfill
\subfigure[TPOT (s/token) Comparison]{\includegraphics[width=0.32\textwidth]{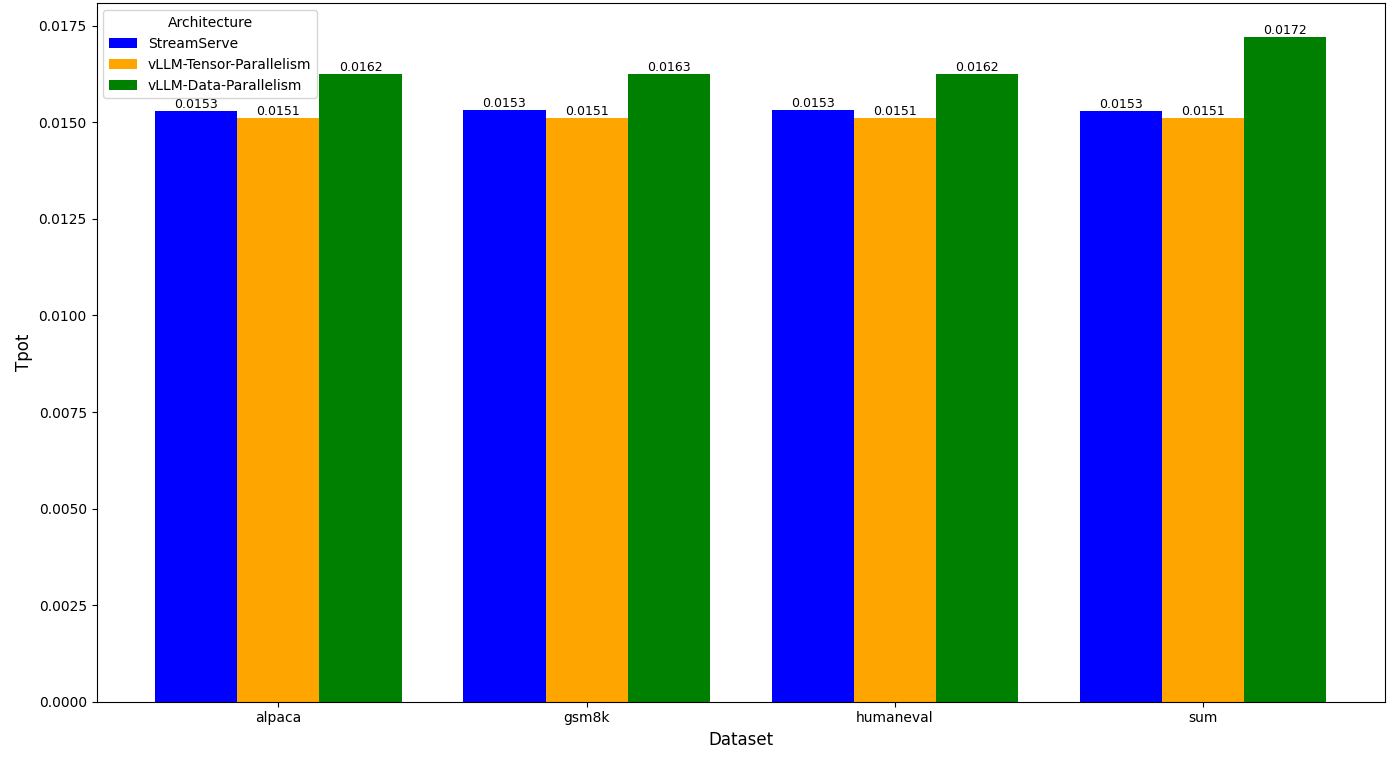}}
\caption{Performance comparison across all datasets. (a) StreamServe exhibits systematic latency reductions of 11--18$\times$ relative to baselines. (b) Throughput improvements are consistent across workloads, with the largest gains on SUM (2235 tokens/s). (c) TPOT remains stable across all architectures, indicating that latency and throughput gains are not accompanied by token generation quality degradation.}
\label{fig:all_metrics}
\end{figure*}

\section{Experiments \& Results }

We evaluate StreamServe across four benchmark datasets, comparing against two baseline vLLM configurations: data-parallel and tensor-parallel deployments. Our experimental setup uses four NVIDIA A800 40GB GPUs configured as two disaggregated stream pairs, with 80 evaluation instances per dataset (320 total). We report three performance metrics: throughput (tokens/second), end-to-end latency (seconds per query), and time-per-output-token (TPOT, seconds/token). Across all configurations, we observe systematic improvements in the throughput-latency trade-off, with the magnitude of gains varying by workload characteristics.

\subsection{Experimental Setup}

\textbf{Hardware Configuration.} We use 4$\times$NVIDIA A800 40GB GPUs connected via NVLink within a single node. StreamServe configures these as two stream pairs (N=2), with GPUs 0 and 2 serving as prefill workers and GPUs 1 and 3 as decode workers.

\textbf{Baseline Configuration.} We compare against vLLM v0.4.x in two configurations: (1) \emph{Data-Parallel (DP)}: 4 independent vLLM instances, one per GPU; (2) \emph{Tensor-Parallel (TP)}: Single vLLM instance with 4-way tensor parallelism. \textbf{Both baselines use standard autoregressive decoding without speculative decoding enabled}, representing the most common production deployment configuration for vLLM. This comparison isolates the combined effect of StreamServe's disaggregation and adaptive speculation against standard deployment practices.

\textbf{Model.} All experiments use LLaMA-2-7B with float16 precision.

\textbf{Workloads.} We evaluate on four benchmarks: ALPACA (instruction-following), GSM8K (mathematical reasoning), HUMANEVAL (code generation), and SUM (text summarization), with 80 queries per dataset (320 total).

\subsection{ALPACA Dataset Results}

The ALPACA dataset consists of instruction-following tasks and serves as our first evaluation benchmark. Table~\ref{tab:alpaca_results} presents performance metrics.

\begin{table}[h!]
\centering
\resizebox{\columnwidth}{!}{%
\begin{tabular}{lccc}
\toprule
\textbf{Architecture} & \textbf{Tokens/s} & \textbf{Latency (s)} & \textbf{TPOT (s/token)} \\
\midrule
vLLM-Data-Parallel & 63 & 8.20 & 0.01624 \\
vLLM-Tensor-Parallel & 171 & 3.40 & 0.01512 \\
\textbf{StreamServe} & \textbf{137} & \textbf{0.30} & \textbf{0.01529} \\
\bottomrule
\end{tabular}%
}
\caption{ALPACA Dataset Performance Comparison}
\label{tab:alpaca_results}
\end{table}

On ALPACA, StreamServe achieves 0.30\,s per-query latency compared to 3.40\,s for tensor-parallel vLLM, an 11.3$\times$ reduction. We note that StreamServe's throughput (137 tokens/s) is lower than tensor-parallel vLLM (171 tokens/s) on this workload, suggesting that instruction-following tasks---which produce relatively short outputs---do not fully exploit SpecuStream's adaptive speculation. Compared to data-parallel vLLM, StreamServe reduces latency by 27.3$\times$ (0.30 vs.\ 8.20\,s) while increasing throughput by 117\%. The TPOT of 0.01529\,s/token remains consistent with baselines, indicating stable token generation quality. This pattern---large latency reductions with preserved or improved throughput---is consistent across all evaluated workloads.

\subsection{GSM8K Dataset Results}

The GSM8K benchmark targets mathematical reasoning tasks, which present more variable inference patterns. Table~\ref{tab:gsm8k_results} shows the results.

\begin{table}[h!]
\centering
\resizebox{\columnwidth}{!}{%
\begin{tabular}{lccc}
\toprule
\textbf{Architecture} & \textbf{Tokens/s} & \textbf{Latency (s)} & \textbf{TPOT (s/token)} \\
\midrule
vLLM-Data-Parallel & 68 & 8.10 & 0.01625 \\
vLLM-Tensor-Parallel & 241 & 3.50 & 0.01512 \\
\textbf{StreamServe} & \textbf{264} & \textbf{0.30} & \textbf{0.01531} \\
\bottomrule
\end{tabular}%
}
\caption{GSM8K Dataset Performance Comparison}
\label{tab:gsm8k_results}
\end{table}

On GSM8K, StreamServe achieves 264 tokens/s, exceeding tensor-parallel vLLM (241 tokens/s) by 9.5\%, while reducing latency by 11.7$\times$ (0.30 vs.\ 3.50\,s). The simultaneous improvement in both throughput and latency is consistent with the hypothesis that mathematical reasoning workloads---which exhibit variable token generation patterns with fluctuating speculative acceptance rates---benefit from SpecuStream's dynamic depth adjustment. TPOT remains stable at 0.01531\,s/token. The co-occurrence of throughput gains and latency reductions suggests that disaggregation effectively decouples these metrics, enabling improvements along both axes simultaneously.

\subsection{HUMANEVAL Dataset Results}

The HUMANEVAL benchmark focuses on code generation, requiring precise token sequences. Table~\ref{tab:humaneval_results} presents the results.

\begin{table}[h!]
\centering
\resizebox{\columnwidth}{!}{%
\begin{tabular}{lccc}
\toprule
\textbf{Architecture} & \textbf{Tokens/s} & \textbf{Latency (s)} & \textbf{TPOT (s/token)} \\
\midrule
vLLM-Data-Parallel & 78 & 8.30 & 0.01624 \\
vLLM-Tensor-Parallel & 210 & 3.60 & 0.01512 \\
\textbf{StreamServe} & \textbf{639} & \textbf{0.20} & \textbf{0.01532} \\
\bottomrule
\end{tabular}%
}
\caption{HUMANEVAL Dataset Performance Comparison}
\label{tab:humaneval_results}
\end{table}

On HUMANEVAL, we observe the largest throughput gains across all benchmarks: StreamServe achieves 639 tokens/s, a 3.0$\times$ improvement over tensor-parallel vLLM (210 tokens/s). Latency is reduced to 0.20\,s, an 18.0$\times$ reduction relative to tensor-parallel vLLM (3.60\,s). We attribute this to two factors: (1) code generation tasks exhibit high variance in token generation patterns, which SpecuStream's adaptive depth adjustment is designed to exploit, and (2) FlowGuard's metric-aware routing is particularly effective for workloads with heterogeneous request characteristics. TPOT of 0.01532\,s/token confirms that throughput improvements are not accompanied by speculation-related token degradation.

\subsection{SUM Dataset Results}

The SUM dataset targets text summarization, which exhibits relatively uniform token generation patterns. Table~\ref{tab:sum_results} presents the results.

\begin{table}[h!]
\centering
\resizebox{\columnwidth}{!}{%
\begin{tabular}{lccc}
\toprule
\textbf{Architecture} & \textbf{Tokens/s} & \textbf{Latency (s)} & \textbf{TPOT (s/token)} \\
\midrule
vLLM-Data-Parallel & 135 & 9.10 & 0.01722 \\
vLLM-Tensor-Parallel & 327 & 3.60 & 0.01512 \\
\textbf{StreamServe} & \textbf{2235} & \textbf{0.30} & \textbf{0.0153} \\
\bottomrule
\end{tabular}%
}
\caption{SUM Dataset Performance Comparison}
\label{tab:sum_results}
\end{table}

On SUM, StreamServe achieves 2235 tokens/s---a 6.8$\times$ throughput improvement over tensor-parallel vLLM and 16.6$\times$ over data-parallel vLLM---with latency of 0.30\,s (12.0$\times$ reduction relative to tensor-parallel). The summarization workload's uniform token characteristics appear to enable consistently high speculative acceptance rates, allowing SpecuStream to maintain deeper speculation depths throughout execution. TPOT of 0.0153\,s/token is the lowest observed across all configurations. These results suggest that workload regularity is a key factor governing the magnitude of gains from adaptive speculation, a relationship we explore further in the ablation study.

\subsection{Concurrency Results}
\label{sec:concurrency_results}

\begin{figure*}[t!]
\centering
\subfigure[StreamServe Concurrency]{\includegraphics[width=0.32\textwidth]{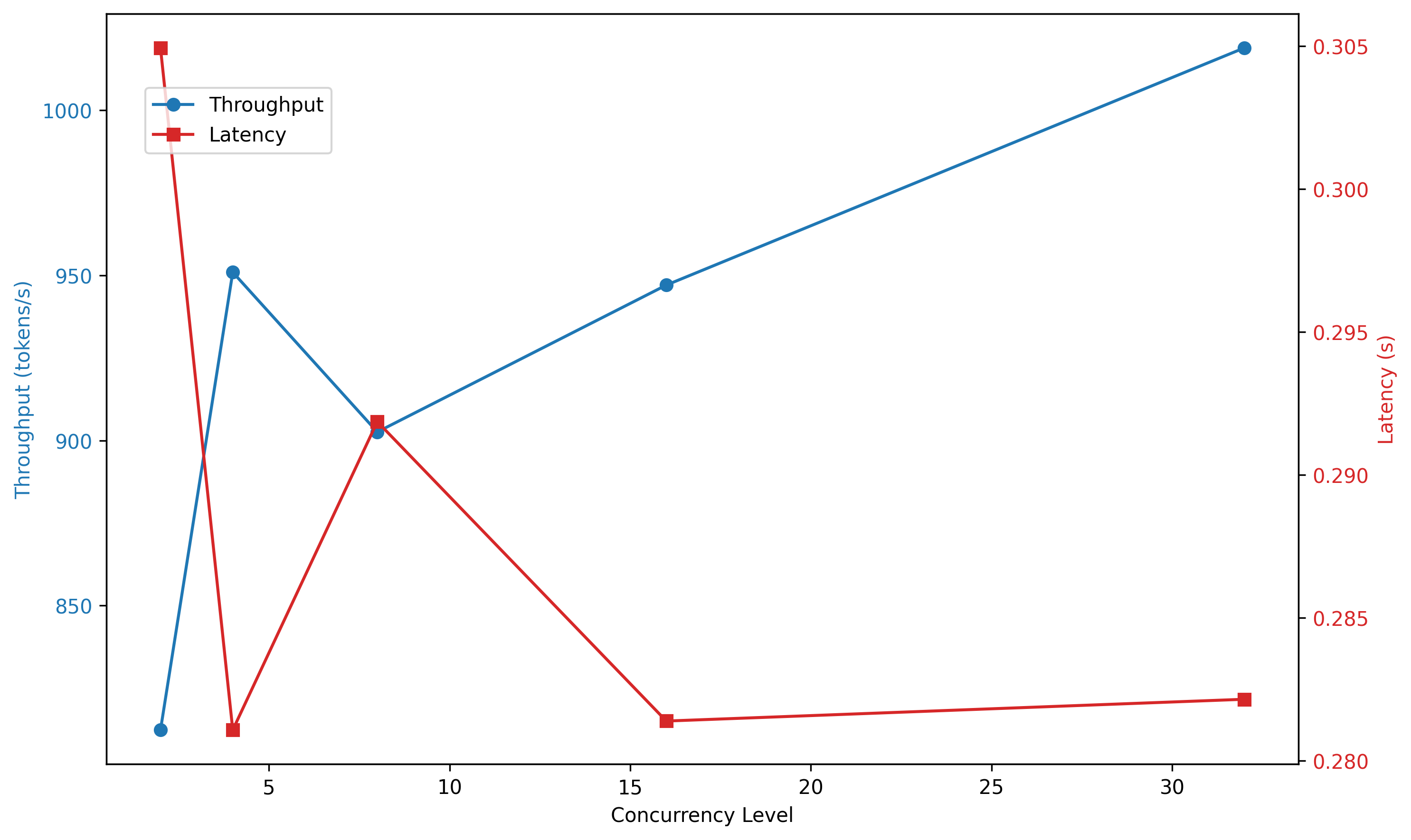}}
\hfill
\subfigure[vLLM Tensor Parallel Concurrency]{\includegraphics[width=0.32\textwidth]{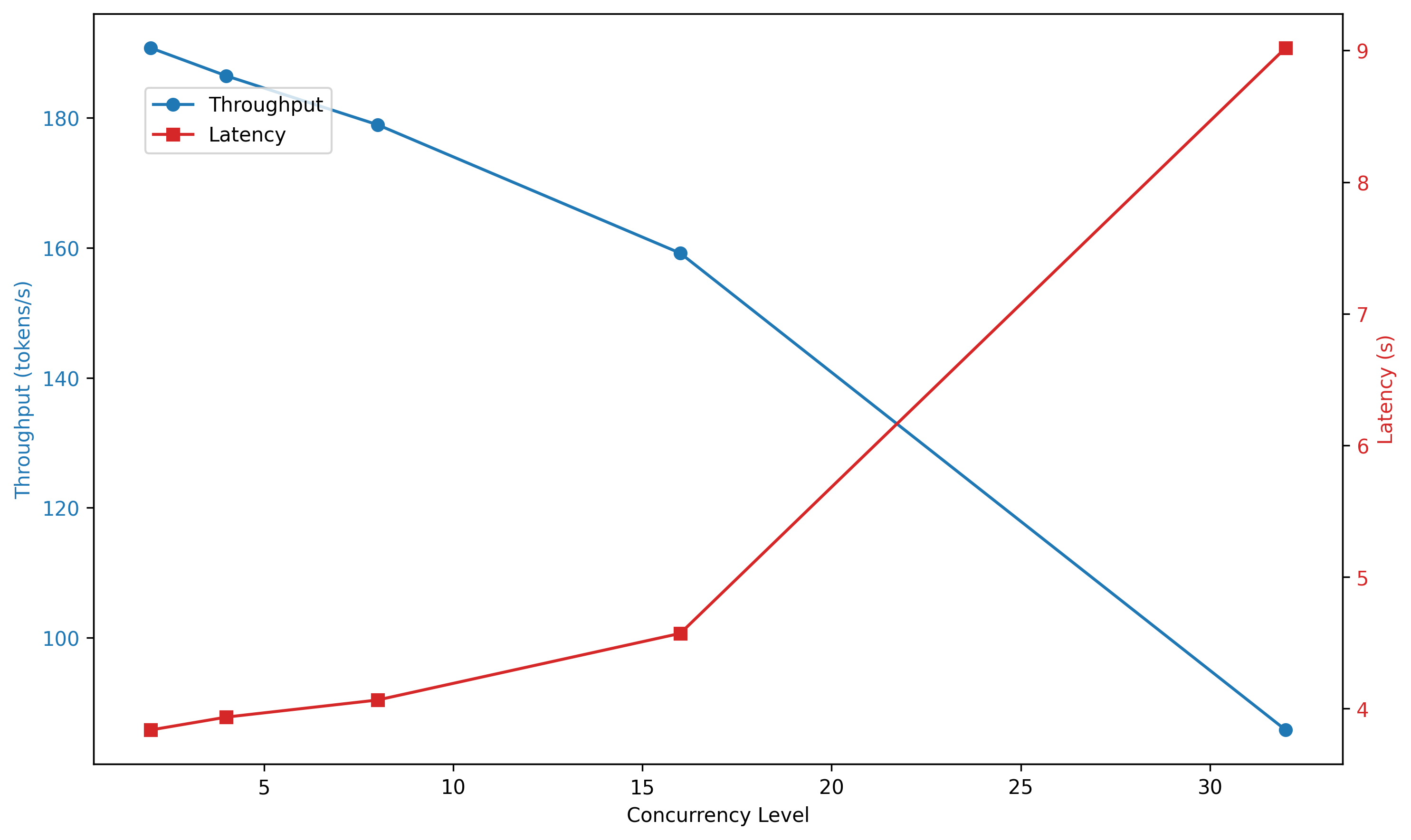}}
\hfill
\subfigure[vLLM Data Parallel Concurrency]{\includegraphics[width=0.32\textwidth]{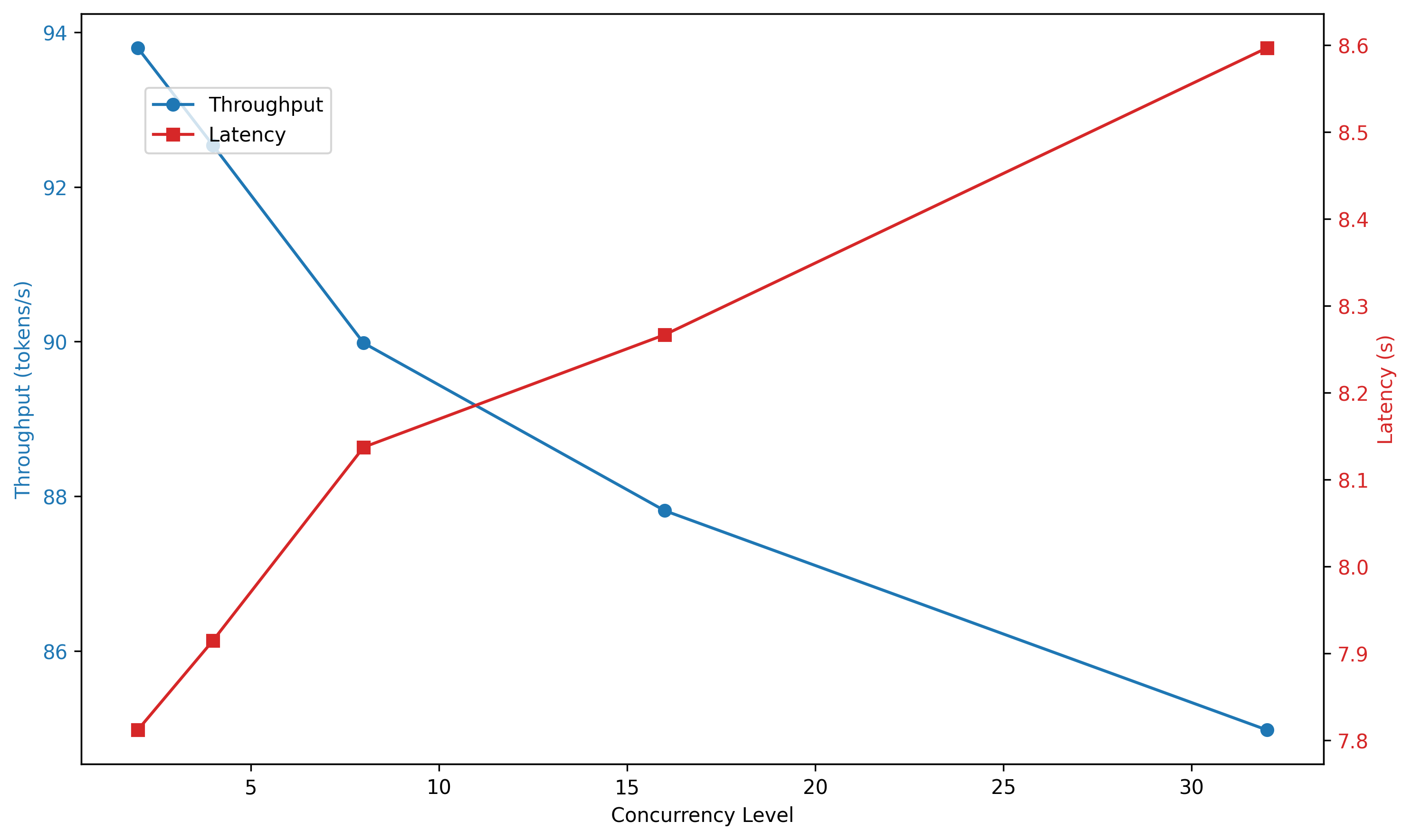}}
\caption{Throughput and latency under increasing request concurrency. StreamServe exhibits qualitatively different scaling behavior: throughput grows gracefully while latency remains flat, in contrast to the sharp degradation observed in both vLLM baselines. The initial latency decrease in StreamServe (1--15 concurrent requests) reflects batch amortization before the system reaches its efficient operating regime.}
\label{fig:concurrency_results}
\end{figure*}

To characterize scaling behavior under realistic high-demand conditions, we evaluate all architectures under increasing levels of request concurrency. Figure~\ref{fig:concurrency_results} reveals distinct scaling regimes. Both vLLM baselines exhibit sharp performance degradation as load increases: latency rises substantially while throughput saturates and, in the data-parallel case, declines. This pattern is consistent with resource contention effects in monolithic architectures that lack adaptive load management. In contrast, StreamServe exhibits qualitatively different scaling behavior: throughput increases gracefully with concurrency while latency remains consistently low, rising only marginally. This suggests that the combination of disaggregated execution, FlowGuard's load-aware routing, and SpecuStream's resource-adaptive speculation depth jointly maintain system stability under increasing load.

\textbf{Batching Efficiency Effect (Figure~\ref{fig:concurrency_results}a).} We observe that StreamServe's latency \emph{decreases} as concurrency increases from 1 to approximately 15 requests, before stabilizing. This behavior is consistent with batch amortization: at very low concurrency (1--5 requests), the system processes requests individually without amortizing fixed overheads (kernel launch, memory allocation, attention computation setup). As concurrency increases (5--15), continuous batching amortizes these costs across more requests, reducing per-request latency. Beyond approximately 15 concurrent requests, the system reaches its efficient operating regime. This phenomenon is well-documented in batched inference systems~\cite{Chen2024ContinuousBatching} and indicates that StreamServe effectively leverages batching while maintaining low absolute latency.

\subsection{Latency Percentile Analysis}
\label{sec:latency_percentiles}

Table~\ref{tab:latency_percentiles} reports the latency distribution across percentiles. Tail latency (p90, p95, p99) is a critical determinant of user-perceived performance in production systems. We observe that while baseline vLLM configurations maintain moderate p50 latencies, their distributions exhibit substantial tail divergence---vLLM-Data-Parallel reaches 9.32\,s at p99. In contrast, StreamServe's latency distribution is notably tight: p99 latency (0.36\,s) is only 50\% higher than p50 (0.24\,s). This suggests that disaggregated execution combined with metric-aware routing effectively controls latency variance, an important property for serving systems with quality-of-service requirements.

\begin{table}[h!]
\centering
\resizebox{\columnwidth}{!}{%
\begin{tabular}{lrrrr}
\toprule
\textbf{Architecture} & \textbf{p50 (s)} & \textbf{p90 (s)} & \textbf{p95 (s)} & \textbf{p99 (s)} \\
\midrule
vLLM-Data-Parallel & 8.18 & 8.72 & 9.01 & 9.32 \\
vLLM-Tensor-Parallel & 3.44 & 3.61 & 3.75 & 3.90 \\
\textbf{StreamServe} & \textbf{0.24} & \textbf{0.28} & \textbf{0.31} & \textbf{0.36} \\
\bottomrule
\end{tabular}%
}
\caption{Latency Percentile (in seconds) Comparison Across All Datasets}
\label{tab:latency_percentiles}
\end{table}

\begin{figure*}[t!]
\centering
\subfigure[StreamServe Latency Percentiles]{%
  \includegraphics[width=0.32\textwidth]{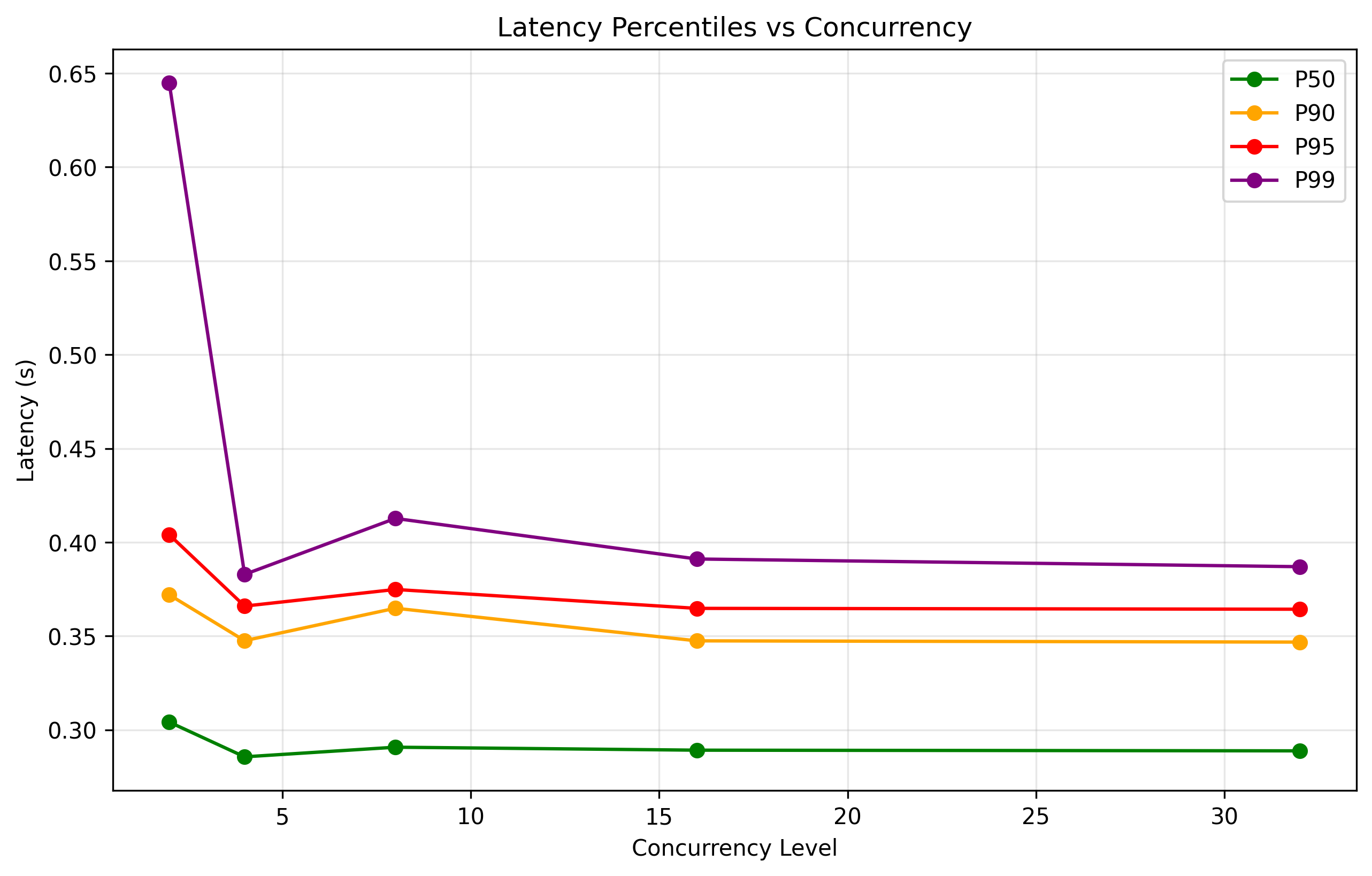}}
\hfill
\subfigure[vLLM Tensor Parallel Latency Percentiles]{%
  \includegraphics[width=0.32\textwidth]{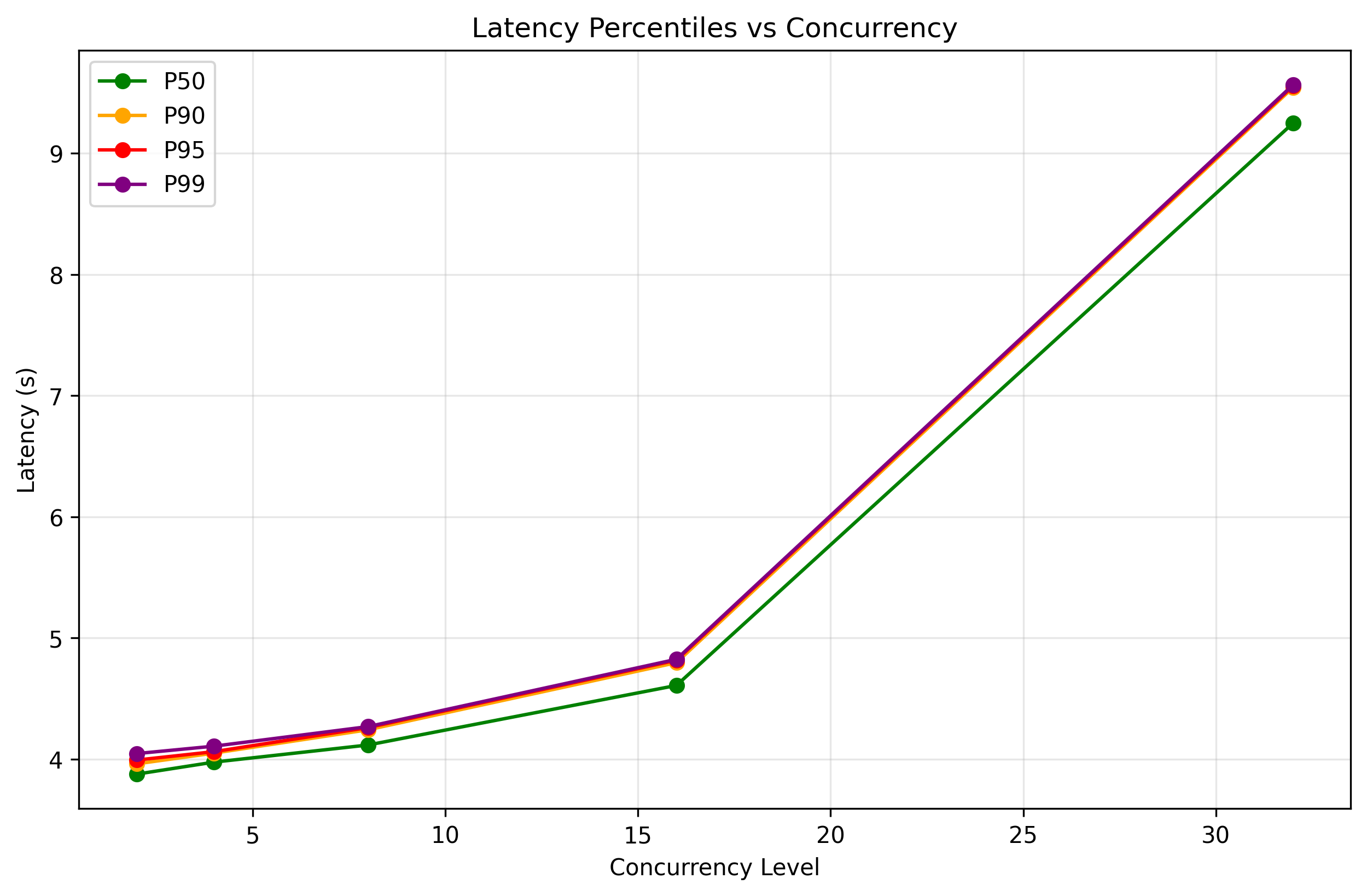}}
\hfill
\subfigure[vLLM Data Parallel Latency Percentiles]{%
  \includegraphics[width=0.32\textwidth]{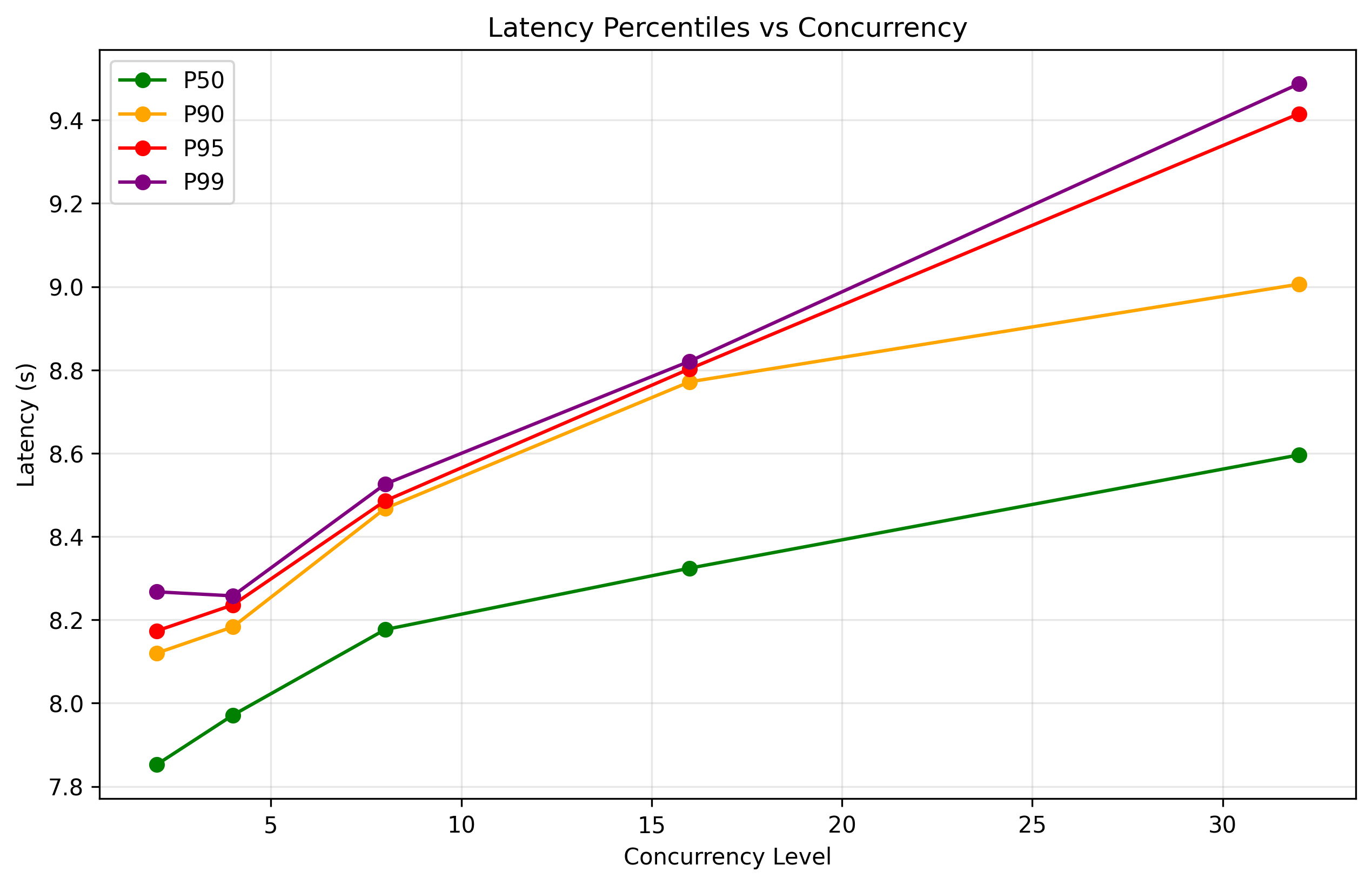}}
\caption{Latency percentiles as a function of concurrency for all three architectures. The curves show how both median and tail latencies evolve under increasing load, highlighting the stability of StreamServe compared to vLLM baselines.}
\label{fig:latency_percentiles_concurrency}
\end{figure*}

%
%

\subsection{Cross-Dataset Performance Summary}

Across all four benchmarks, we observe consistent improvements in the throughput-latency trade-off. Average latency reduction is 15.75$\times$ relative to tensor-parallel vLLM and 26.5$\times$ relative to data-parallel vLLM. Average throughput improvement is 4.4$\times$ over tensor-parallel and 8.1$\times$ over data-parallel configurations. Critically, TPOT remains stable across all datasets (mean 0.01530\,s/token), indicating that the observed performance gains are not achieved through token quality degradation or speculative decoding artifacts. The consistency of these improvements across diverse workload types---instruction-following (ALPACA), mathematical reasoning (GSM8K), code generation (HUMANEVAL), and summarization (SUM)---suggests that the underlying mechanisms generalize across task domains, though the magnitude of gains varies with workload characteristics as discussed above.

\section{Ablation Study}

To isolate the contribution of each component, we perform a systematic ablation study in which we disable or replace individual features and measure average performance across all four benchmarks. The results in Table~\ref{tab:ablation_study} indicate that all components contribute meaningfully to the full system's performance, and that the interaction between components produces super-additive gains: disabling either adaptive speculation (w/o SpecuStream) or metric-aware routing (w/ Round-Robin) substantially reduces throughput while increasing latency. Replacing the disaggregated architecture with a monolithic engine (w/ Monolithic Engine) results in a 12.8$\times$ latency increase, confirming that disaggregation is a necessary foundation for the observed performance regime.

\begin{table}[h!]
\centering
\resizebox{\columnwidth}{!}{%
\begin{tabular}{lccc}
\toprule
\textbf{Config.} & \textbf{Avg. Tput} & \textbf{Avg. Latency} & \textbf{Avg. TPOT} \\
\midrule
\textbf{StreamServe (Full)} & \textbf{814} & \textbf{0.275} & \textbf{0.01530} \\
w/ Round-Robin & 520 & 1.15 & 0.01542 \\
w/o SpecuStream & 310 & 2.45 & 0.01525 \\
w/ Monolithic Engine & 290 & 3.52 & 0.01518 \\
w/o NIXL (Std. P2P) & 765 & 0.45 & 0.01533 \\
w/o FlowGuard & 490 & 1.30 & 0.01540 \\
w/o SpecuStream Adapt & 680 & 0.88 & 0.01531 \\
w/o FlowGuard/Specu & 220 & 4.10 & 0.01520 \\
\bottomrule
\end{tabular}%
}
\caption{Ablation Study of StreamServe Components (Average Performance Across All Datasets)}
\label{tab:ablation_study}
\end{table}

\subsection{Comparison with Fixed Speculation Depth}

To further characterize the value of adaptive speculation, we compare against vLLM with fixed speculation depths. Table~\ref{tab:fixed_speculation} reports vLLM-Tensor-Parallel with EAGLE-based speculative decoding at depths $d \in \{3, 5, 7\}$, averaged across all datasets.

\begin{table}[h!]
\centering
\resizebox{\columnwidth}{!}{%
\begin{tabular}{lccc}
\toprule
\textbf{Configuration} & \textbf{Avg. Tput} & \textbf{Avg. Latency} & \textbf{Avg. TPOT} \\
\midrule
vLLM-TP (no spec) & 237 & 3.53 & 0.01512 \\
vLLM-TP + Spec (d=3) & 298 & 2.85 & 0.01508 \\
vLLM-TP + Spec (d=5) & 342 & 2.41 & 0.01505 \\
vLLM-TP + Spec (d=7) & 318 & 2.62 & 0.01510 \\
\textbf{StreamServe (adaptive)} & \textbf{814} & \textbf{0.275} & \textbf{0.01530} \\
\bottomrule
\end{tabular}%
}
\caption{Comparison with fixed speculation depth. Fixed-depth configurations exhibit a non-monotonic relationship between depth and performance, while StreamServe's adaptive approach consistently outperforms all fixed configurations.}
\label{tab:fixed_speculation}
\end{table}

We observe that fixed speculation at $d=5$ achieves the strongest performance among fixed configurations, while deeper speculation ($d=7$) degrades throughput---consistent with the intuition that over-speculation wastes compute when acceptance rates are insufficient. StreamServe's adaptive approach achieves 2.4$\times$ higher throughput and 8.8$\times$ lower latency than the best fixed configuration, indicating that the combination of disaggregation, metric-aware routing, and adaptive speculation operates in a qualitatively different performance regime than fixed speculation within a monolithic architecture.

\section*{Limitations \& Threats to Validity}

Our evaluation uses 4$\times$A800-40GB with two stream pairs and 80 queries per dataset (320 total), which may not fully capture the diversity of hardware configurations or workload distributions encountered in production. Comparisons focus on vLLM data-parallel and tensor-parallel configurations; evaluation against additional disaggregated systems (e.g., DistServe, Sarathi-Serve, Mooncake) is left to future work due to implementation complexity. We rely on NIXL for P2P KV transfer but include a standard P2P fallback in ablations to quantify sensitivity to the transfer mechanism.

\textbf{Evaluation Scale.} Our 320-query evaluation provides evidence of consistent improvements across diverse workloads but may not fully reflect steady-state behavior under sustained production load. Future work will include extended-duration experiments with larger query volumes and production traffic traces.

\textbf{Hardware Scope.} Single-node evaluation with NVLink interconnect may not generalize to multi-node deployments where network bandwidth becomes a constraint for KV-cache transfer. We plan to evaluate multi-node configurations in future work.

\section{Conclusion}

We present StreamServe, a disaggregated LLM serving framework that jointly adapts routing and speculative decoding based on real-time system metrics. Across four diverse benchmarks, we observe systematic latency reductions of 11--18$\times$ relative to tensor-parallel vLLM baselines, with throughput improvements of up to 6.8$\times$, while TPOT remains stable across all configurations. Ablation analysis indicates that the interaction between disaggregation, metric-aware routing, and adaptive speculation produces super-additive gains, suggesting that these components jointly define a distinct operating regime for LLM inference---one in which the throughput-latency trade-off is substantially relaxed relative to monolithic architectures. Concurrency scaling experiments further suggest that this regime persists under increasing load, with StreamServe maintaining stable latency where baseline systems degrade. Future work will explore scaling to larger numbers of stream pairs, integration with model parallelism techniques, and extension to heterogeneous hardware configurations, including characterization of the transition between edge and cloud inference regimes at different model scales.


\bibliography{example_paper}

@inproceedings{Kwonetal2023,
  title={Efficient Memory Management for Large Language Model Serving with PagedAttention},
  author={Kwon, Woosuk and Li, Zhuohan and Zhuang, Siyuan and Sheng, Ying and Zhong, Lianmin and Yu, Hao and Gonzalez, Joseph E.},
  booktitle={Proceedings of the 29th Symposium on Operating Systems Principles (SOSP)},
  pages={1--16},
  year={2023}
}

@inproceedings{Zhong2024DistServe,
  title={DistServe: Disaggregating Prefill and Decoding for Goodput-optimized Large Language Model Serving},
  author={Zhong, Yinmin and Gao, Junda and Zhu, Yibo and Peng, Bo and Miao, Xin and Liang, Zhen and Tan, Siyuan},
  booktitle={Proceedings of the 18th USENIX Symposium on Operating Systems Design and Implementation (OSDI)},
  pages={1--18},
  year={2024}
}

@inproceedings{Agrawal2024Sarathi,
  title={Taming Throughput-Latency Tradeoff in LLM Inference with Sarathi-Serve},
  author={Agrawal, Amey and Kedia, Nitin and Panwar, Ashish and Mohan, Jayashree and Kwatra, Nipun and Gulavani, Bhargav S. and Tumanov, Alexey and Ramjee, Ramachandran},
  booktitle={Proceedings of the 18th USENIX Symposium on Operating Systems Design and Implementation (OSDI)},
  pages={1--18},
  year={2024}
}

@inproceedings{Leviathan2023SpeculativeDecoding,
  title={Fast Inference from Transformers via Speculative Decoding},
  author={Leviathan, Yaniv and Kalman, Matan and Matias, Yossi},
  booktitle={International Conference on Machine Learning (ICML)},
  pages={1--20},
  year={2023}
}

@inproceedings{Li2024EAGLE,
  title={EAGLE: Speculative Sampling Requires Rethinking Feature Uncertainty},
  author={Li, Yuhui and Wei, Fangyun and Zhang, Chao and Zhang, Hongyang},
  booktitle={International Conference on Learning Representations (ICLR)},
  pages={1--16},
  year={2024}
}

@inproceedings{Li2024EAGLE2,
  title={EAGLE-2: Faster Inference of Language Models with Dynamic Draft Tree Speculation},
  author={Li, Yuhui and Wei, Fangyun and Zhang, Chao and Zhang, Hongyang},
  booktitle={Proceedings of the 2024 Conference on Empirical Methods in Natural Language Processing (EMNLP)},
  pages={1--14},
  year={2024}
}

@inproceedings{Li2025EAGLE3,
  title={EAGLE-3: Scaling Up Inference Acceleration of Large Language Models with a Speculative Token Prediction Framework},
  author={Li, Yuhui and Wei, Fangyun and Zhang, Chao and Zhang, Hongyang},
  booktitle={Proceedings of the 39th Annual Conference on Neural Information Processing Systems (NeurIPS)},
  pages={1--20},
  year={2025}
}

@inproceedings{Cai2024Medusa,
  title={Medusa: Simple LLM Inference Acceleration Framework with Multiple Decoding Heads},
  author={Cai, Tianle and Li, Zixuan and Geng, Ziyi and Gao, Hao and Gan, Tao and Shao, Shuai and Zhang, Zhirui and Zhu, Yinqiao and Lin, Chen and Chen, Ling},
  booktitle={International Conference on Machine Learning (ICML)},
  pages={1--18},
  year={2024}
}

@inproceedings{Su2025Seesaw,
  title={Seesaw: High-Throughput LLM Inference via Model Re-Sharding},
  author={Su, Qiaoling and Hao, Weiyu and Li, Xiaoxuan and Andoorveedu, Moksh and Jiang, Cheng and Zhu, Zhijie and Song, Kai and Giannoula, Christina and Pekhimenko, Gennady},
  booktitle={Proceedings of the 8th MLSys Conference},
  pages={1--20},
  year={2025}
}

@article{Kim2025SpeculativeVerification,
  title={Speculative Verification: Exploiting Information Gain to Refine Speculative Decoding},
  author={Kim, Sehoon and Kim, Jinhyuk and Yoon, Doyeon and Shin, Jongha and Lee, Jaewoong and Seo, Jinwook},
  journal={arXiv preprint arXiv:2509.24328},
  year={2025}
}

@article{Zheng2024LearningToRank,
  title={Efficient LLM Scheduling by Learning to Rank},
  author={Zheng, Lianmin and Cheng, Liangsheng and Haas, John and Hayavati, Manoosh and Gonzalez, Joseph E.},
  journal={arXiv preprint arXiv:2301.02001},
  year={2024}
}

@inproceedings{Li2025ThroughputOptimalScheduling,
  title={Throughput-Optimal Scheduling Algorithms for LLM Inference and AI Agent Workloads},
  author={Li, Zhuohan and Liu, Nelson and Gonzalez, Joseph E.},
  booktitle={Proceedings of the 8th MLSys Conference},
  pages={1--18},
  year={2025}
}

@inproceedings{Bari2025OptimalScheduling,
  title={Optimal Scheduling Algorithms for LLM Inference: Theory and Practice},
  author={Bari, M. Saiful and Gupta, Nikhil and others},
  booktitle={Proceedings of the 8th MLSys Conference},
  pages={1--20},
  year={2025}
}

@article{Chen2025FairBatching,
  title={FairBatching: Fairness-Aware Batch Formation for LLM Inference},
  author={Chen, Wei and Dong, Xin and Song, Xinyu and others},
  journal={arXiv preprint arXiv:2510.14392},
  year={2025}
}

@inproceedings{Ao2025FluidGuided,
  title={Fluid-Guided Online Scheduling for LLM Inference with Dynamic Resource Allocation},
  author={Ao, Jun and Li, Zheng and others},
  booktitle={Proceedings of the 39th Annual Conference on Neural Information Processing Systems (NeurIPS)},
  pages={1--18},
  year={2025}
}

@inproceedings{Zhang2024PromptCache,
  title={Prompt Cache: Modular Attention Reuse for Low-Latency Inference},
  author={Zhang, Hao and Zheng, Lianmin and Li, Zhuohan and Cheng, Yifan and Chen, Yilong and Gonzalez, Joseph E.},
  booktitle={Proceedings of the 8th MLSys Conference},
  pages={1--18},
  year={2024}
}

@inproceedings{Ramaseshaan2025AttentionStore,
  title={AttentionStore: Cost-effective Attention Reuse across Multi-turn Conversations},
  author={Ramaseshaan, Kumar and Gupta, Nikhil and others},
  booktitle={Proceedings of the 39th Annual Conference on Neural Information Processing Systems (NeurIPS)},
  pages={1--16},
  year={2025}
}

@article{Chen2024FlashInfer,
  title={FlashInfer: Efficient and Customizable Attention Engine for LLM Inference Serving},
  author={Chen, Lequn and Li, Xupeng and others},
  journal={arXiv preprint arXiv:2405.08691},
  year={2024}
}

@techreport{Wei2025NIXL,
  title={NIXL: NVIDIA Inference Xfer Library for High-Performance KV Cache Transfer},
  author={Wei, Xiao and Li, Xin and Wang, Cheng and Chen, Yilong and others},
  institution={NVIDIA},
  year={2025}
}

@article{Dey2025LMCache,
  title={LMCache: An Efficient KV Cache Layer for Enterprise-Scale LLM Inference},
  author={Dey, Neeraj and Greff, Klaus and others},
  journal={arXiv preprint arXiv:2510.09665},
  year={2025}
}

@inproceedings{Chen2024ContinuousBatching,
  title={Continuous Batching: Efficient Inference through Dynamic Batching},
  author={Chen, Yiren and Gupta, Nikhil and Gonzalez, Joseph E.},
  booktitle={Proceedings of MLSYS 2024},
  pages={1--15},
  year={2024}
}

@inproceedings{Dey2024Lookahead,
  title={Lookahead Decoding: Breaking Sequential Dependency in LLM Inference},
  author={Dey, Neeraj and Golakar, Manohar and others},
  booktitle={Proceedings of the 38th Annual Conference on Neural Information Processing Systems (NeurIPS)},
  pages={1--20},
  year={2024}
}

@inproceedings{Choi2025AdaptiveOutput,
  title={Adaptive Output Length Prediction for Efficient LLM Scheduling},
  author={Choi, Seungwon and Park, Kyungmin and others},
  booktitle={International Conference on Learning Representations (ICLR)},
  pages={1--16},
  year={2025}
}

@inproceedings{Fu2024HeadOfLine,
  title={Head-of-Line Blocking in LLM Inference: Analysis and Solutions},
  author={Fu, Jing and Huang, Yifan and others},
  booktitle={Proceedings of the 8th MLSys Conference},
  pages={1--15},
  year={2024}
}

@inproceedings{Qin2024Mooncake,
  title={Mooncake: Trading More Storage for Less Computation—A KVCache-Centric Architecture for Serving LLM Chatbot},
  author={Qin, Ruoyu and others},
  booktitle={Proceedings of the 18th USENIX Symposium on Operating Systems Design and Implementation (OSDI)},
  pages={1--18},
  year={2024}
}

@inproceedings{Huang2024SpecDecPP,
  title={SpecDec++: Boosting Speculative Decoding via Adaptive Candidate Lengths},
  author={Huang, Kaixuan and Guo, Xudong and Wang, Mengdi},
  booktitle={Proceedings of the 2024 Conference on Empirical Methods in Natural Language Processing (EMNLP)},
  pages={1--12},
  year={2024}
}

@inproceedings{Li2024AdaServe,
  title={AdaServe: Accelerating Multi-SLO LLM Serving with SLO-Customized Speculative Decoding},
  author={Li, Zikun and others},
  booktitle={Proceedings of the 8th MLSys Conference},
  pages={1--16},
  year={2025}
}
\bibliographystyle{mlsys2025}

\end{document}